\documentclass{elsart}
\def\ba{\begin{eqnarray}}
\def\ea{\end{eqnarray}}
\input{epsf.tex}

\usepackage{amssymb}

\begin{document}
\date{}
\begin{frontmatter}



\title{MERADGEN 1.0: Monte Carlo generator for the simulation of radiative
events in parity conserving doubly-polarized M{\o}ller scattering}


\author[JLab]{Andrei Afanasev}
\author[JLab]{Eugene Chudakov}
\author[Minsk]{Alexander Ilyichev}
\ead{ily@hep.by}
\author[Gomel]{Vladimir Zykunov}
\ead{zykunov@sunse.jinr.ru}

\address[JLab]{Jefferson Lab, Newport News, VA 23606, USA}
\address[Minsk]{National Scientific
and Educational Centre of Particle and High Energy Physics 
of the Belarusian State University, Minsk, 220040  Belarus
}
\address[Gomel]{Joint Institute for Nuclear Research, Dubna, 141980 Russia 
and
Gomel State Technical University, Gomel, 246746 Belarus}

\begin{abstract}
The Monte Carlo generator MERADGEN 1.0 for the simulation of radiative
events in parity conserving doubly-polarized M{\o}ller scattering 
has been developed.
Analytical integration wherever it is possible provides rather fast
and accurate generation. 
Some numerical tests and histograms are presented.
\end{abstract}

\end{frontmatter}

\section*{PROGRAM SUMMARY}
{\it Manuscript title:}
MERADGEN 1.0: Monte Carlo generator for the simulation of radiative
events in parity conserving doubly-polarized M{\o}ller scattering\\
{\it Authors:} A. Afanasev, E. Chudakov, A. Ilyichev, V. Zykunov\\
{\it Program title:} MERADGEN 1.0\\
{\it Licensing provisions:} none\\
{\it Programming language:} FORTRAN 77\\
{\it Computer(s) for which the program has been designed:} all\\
{\it Operating system(s) for which the program has been designed:} Linux\\
{\it RAM required to execute with typical data:} 1 MB\\
{\it Has the code been vectorised or parallelized?:} no\\
{\it Number of processors used:} 1\\
{\it Supplementary material:} none\\
{\it Keywords:} radiative corrections, Monte Carlo method,
M{\o}ller Scattering\\
{\it \underline{PACS}:} 07.05.Tp, 13.40.Ks, 13.88.+e, 25.30.Bf\\
{\it \underline{CPC Library Classification}:} \\
{\it External routines/libraries used:} none\\
{\it CPC Program Library subprograms used:}  none\\

\ \\
{\it Nature of problem:} simulation of radiative
events in parity conserving doubly-polarized M{\o}ller scattering.\\
{\it Solution method:} Monte Carlo method for simulation within QED,
analytical integration wherever it is possible that 
provides rather fast and accurate generation\\
{\it Restrictions:} none\\
{\it Unusual features:} none\\
{\it Additional comments:} none\\
{\it Running time:} the simulation of $10^8$ radiative events 
for $itest:=1$ takes up to 45 seconds on AMD Athlon 2.80 GHz
processor.

\section{Introduction}
The precise measurements in polarized M{\o}ller scattering play
a very important role in the modern polarimetry. The coincidence detection
of the final electron pairs allows to essentially reduce a background of
radiative effects \cite{asym} that accompany any charge particle scattering,
including investigated reaction. 
However due to finite detector resolution it is impossible to remove 
all radiative event contributions out of the data. 
Moreover the additional virtual
particle contributions can not be removed by any experimental cuts.
Therefore to reach the appropriate accuracy
we need to perform the radiative correction procedure  
consisting in the calculation of these effects within QED 
and estimate them numerically.

The calculation of the lowest order QED radiative corrections (RC) 
to polarized M{\o}ller scattering was already performed in 
\cite{suarez,mera} (see also the references therein). 
In \cite{suarez} these corrections were
calculated exactly (without the ultrarelativistic approximation) 
but unfortunately also without any experimental cuts.
In \cite{mera} the similar corrections were performed
for longitudinally polarized M{\o}ller scattering 
within the ultrarelativistic approximation. 
The numerical analysis which is presented in \cite{mera} shows 
that RC to M{\o}ller scattering are very sensitive 
to the missing mass (inelasticity) cuts.

However, the consideration only the missing mass cuts during RC procedure 
is insufficient. The realistic situation corresponds to taking into account 
the detector geometry that can essentially complicate the integration over 
the real photon phase space. 
In such  situation an approach based on the Monte Carlo simulation
of radiative events is the most adequate. The Monte Carlo generators RADGEN 
\cite{RAD} and ELRADGEN \cite{ELRAD} for simulation of radiative events
in deep inelastic and elastic lepton-nucleon scattering
can serve as an examples of such approach.   

In present paper a new Monte Carlo generator MERADGEN 1.0
\footnote{FORTRAN code MERADGEN 1.0 is available from http://www.hep.by/RC}
for the simulation of the radiative events in parity conserving part of 
polarized M{\o}ller scattering is presented
that can be used for the polarimetry purpose of present and future 
experiments, for example, in JLab and SLAC. 
Naturally, as we are taking into consideration the parity conserving 
effects and the beam energy is rather small (from 1 to 45 GeV),
the weak contributions ($Z$-boson exchange and so one) 
to this process are negligible.
So, we restricted our calculation within QED theory only.   

The present paper is organized as following. 
In Section \ref{kin} the kinematics
of the investigated process and the generation method are presented.
The different contributions to the cross section that is responsible
for real photon emission are considered in Section \ref{xxsect}. 
Then in Section \ref{sstruct}
the brief structure of the code is discussed, in Section 5 we described
the input-output data. In Section \ref{ttest} we explain
how to run tests of our code, some conclusions are given 
in Section \ref{cconc}. Finally, in Appendices the 
4-momenta reconstruction formulas, some lengthy formulas
for RC and test output are presented.

\section{Kinematics and Method of Generation}
\label{kin}
The lowest order contribution to the observable cross section 
of M{\o }ller scattering reaction 
\ba
e(k_1,\xi_L)+e(p_1,\eta_L)\to
e(k_2)+e(p_2)
\label{el}
\ea
(in parentheses the 4-momenta and polarization vectors
of electrons are presented, and $k_1^2=k^2_2=p_1^2=p_2^2=m^2$)
can be  described by the standard set of Mandelstam variables:
\ba
&&s=(k_1+p_1)^2,\ t=(k_1-k_2)^2,\ u=(k_2-p_1)^2,
\nonumber \\ 
&&s+t+u=4 m^2,
\label{m1}
\ea 
while the beam ($\xi_L$) and target  ($\eta_L$) 
polarization vectors read:
\ba
\xi_L=
\frac 1{\sqrt{\lambda _s}}
\Bigl(\frac{s-2 m^2}mk_1-2 mp_1 \Bigr),
\nonumber \\
\eta_L=
\frac 1{\sqrt{\lambda _s}}
\Bigl(2 mk_1-\frac{s-2 m^2}mp_1 \Bigr).
\ea  
where $\lambda _s=s(s-4m^2)$.

For the definition of the lowest order contribution to
M{\o}ller scattering it is enough to define an initial 
beam energy $k_{10}=E_b^{\rm Lab} $ (in Lab. system), 
a scattering angle $\theta _{\rm CM}$ (in CM system)
of the detected electron with the 4-momentum $k_2$ and an azimuthal angle
$\phi $. The cross section does not depend on $\phi $
up to taking into account the detector geometry. 
Let us notice that $E^{\rm Lab}_b$ and 
$\cos \theta _{\rm CM}$
can be expressed via the invariants in the following way:
\ba
\cos \theta_{\rm CM}=1+2t/s=1-2y,\;
E_{b}^{\rm Lab}=\frac{s-2m^2}{2m},
\label{t0}
\ea
where $y=-t/s$.
Therefore instead of the scattering angle and the beam energy 
for the definition of the Born cross section
we can use $s$ and $t$ variables.

For the radiative process with the real photon emission
\ba
e(k_1,\xi_L)+e(p_1,\eta_L)\to
e(k_2)+e(p_2)+\gamma(k)
\label{in}
\ea
($k^2=0$) three new kinematic variables have to be defined.
At first, following notations of ref. \cite{mera} we introduce an inelasticity
$v=\Lambda^2-m^2$, where $\Lambda=k_1-k_2+p_1$ and
$\Lambda^2$ is so-called missing mass square. 
Maximum value of the inelasticity
\ba
v_{max}=\frac{st+\sqrt{s(s-4m^2)t(t-4m^2)}}{2m^2}\sim s+t 
\label{vm}
\ea
can be defined from the kinematical restriction
(see, for example, the Chew-Low diagram in \cite{zyk}).
This variable can be directly reconstructed from the data.
To remove the contribution of the hard photon emission 
the events with $v\le v_{cut}$ are taking into account, where 
the value $v_{cut}$ is far less than $v_{max}$.
The second variable is defined by
$t_1=(p_2-p_1)^2=(k_1-k_2-k)^2$.
At last, the third variable should be choose as
$z=2k_2k$. Notice that for the radiative process
\begin{equation}
\cos \theta_{\rm CM}=1+2t/(s-v).
\label{tr}
\end{equation}

To reconstruct the 4-momenta of all
particles for radiative process in any system
it is enough to determine of variables $s$, $t$, $v$, $t_1$, $z$
and the azimuthal angle $\phi $.  
As an example, in \ref{av} the 4-momenta of electrons and real photon
are expressed through these variables and presented in the center 
of the initial electron mass system.   

The simulation of the radiative events 
can be performed by the following algorithm:
\begin{itemize}
\item For the fixed initial energy and $t$ the non-radiative
and radiative parts of the observable cross section are calculated.
\item The channel of scattering is simulated for the given event
in accordance with partial contributions of these two
(non-radiative and radiative) positive parts
into the observable cross section.
\item The angle $\phi$ is simulated uniformly from $0$ to $2\pi$.
\item For the radiative event the kinematic variables $v$, $t_1$ and $z$
are simulated in accordance with their calculated distributions.
\item The 4-momenta of all final particles in the required
system are calculated.
\item If the initial $t$ has not a fixed value (i.e. simulated
according to the Born probability distribution) 
then the cross sections have to be stored for reweighing.
The  $t$-distribution is simulated over the Born cross section,
and realistic observed $t$-distribution is calculated as sum of
weights, they are ratios of the observable and Born cross sections.
\end{itemize}

Let us consider some important steps of simulation  of the radiative events 
in more details.

\section{Non-radiative and radiative parts of the
observable cross section}
\label{xxsect}
Here we consider 
the observable cross section that has a form
\footnote{Here and later we consider
the differential cross section $\sigma\equiv d\sigma/dy$ }:
\ba
\sigma^{obs}(v_{cut})
=\sigma^0+\sigma^{RV}(v_{cut})+\sigma^{R}_F(v_{cut}),
\label{st}
\ea
where $\sigma^0$ is the Born contribution,
$\sigma^{RV}$ is an infrared divergency free sum of
the contributions of the additional virtual particles 
and the "infrared" part  of the real photon emission,
$\sigma^{R}_F$ is the infrared  divergency free part 
of the unobservable photon emission.
The explicit expressions for each term of equation (\ref{st})
can be found in \cite{mera}. 
Since the expressions for the two virtual photon 
contribution as well as $\delta _1^H $ and $\delta _1^S $ 
in \cite{mera} contain the misprints (fortunately, it is not seriously
reflecting on the numerical estimations
presented in \cite{mera}), the additional virtual particle 
contributions together with $\delta _1^H $ and $\delta _1^S $ 
are also presented in \ref{avc}. 

Now we consider $\sigma^{R}_F$ in more detail:
before integration over the inelasticity and
the real photon phase space
it can be presented as
\ba
\sigma^R_{F}(v_{cut})=
\frac{\alpha^3}{4s}\int\limits_{0}^{v_{cut}}
dv\int d\Gamma _k (|M_R|^2-4F^{IR}|M_0|^2),
\label{rf}
\ea
where $M_R$ is a sum of matrix elements contributed to 
the real photon emission (see \ref{arc} for details), 
while $M_0$ is a Born matrix element.
The real photon phase space reads
\ba
d\Gamma _k =\frac 1{\pi}
\frac {d^3k}{k_0}\delta ((\Lambda-k)^2-m^2)
=\frac 1{4\pi}\frac {dt_1dz }
{\sqrt{-\Delta (k_1,k_2,p_1,k)}},
\ea
where $\Delta (k_1,k_2,p_1,k)$ is the Gram determinant.
At last,  
\ba
-F^{IR}&=&
\frac {m^2}{z^2}+
\frac {m^2}{z_1^2}+
\frac {m^2}{v^2}+
\frac {m^2}{v_1^2}+
\frac {s-2m^2}{z_1v_1}+
\frac {s-2m^2}{zv}+
\nonumber \\  &&
+\frac {t-2m^2}{z_1z}+
\frac {t-2m^2}{v_1v}+
\frac {u_0-2m^2}{z_1v}+
\frac {u_0-2m^2}{v_1z},
\ea
where $z_1=2kk_1$, $v_1=2kp_1$ and $u_0=s+t-4m^2$.

For separation of the cross section~(\ref{st}) into the radiative 
and non-radiative parts it is necessary to introduce a new fictitious  
parameter $v_{min}$ associated with missing mass square resolution.
Then the equation (\ref{rf})
can be rewrite in the following way
\ba
\sigma^R_{F}(v_{cut})&=&\sigma^{r}(v_{cut},v_{min})+
\frac{\alpha^3}{4s}\int\limits_{0}^{v_{min}}
dv\int d\Gamma _k (|M_R|^2-4F^{IR}|M_0|^2) -
\nonumber \\&&
-
\frac{\alpha^3}{s}\int\limits_{v_{min}}^{v_{cut}}
dv\int d\Gamma _k F^{IR}|M_0|^2,
\label{rf2}
\ea
where
\ba
\sigma^{r}(v_{cut},v_{min})&=&
\frac{\alpha^3}{16\pi s}\int\limits_{v_{min}}^{v_{cut}}
dv\int\limits_{t_1^{min}}^{t_1^{max}}dt_1
\int\limits_{z^{min}}^{z^{max}}
\frac{dz}{\sqrt{-\Delta (k_1,k_2,p_1,k)}}
  |M_R|^2 =
\nonumber \\&=&
\int\limits_{v_{min}}^{v_{cut}}
dv\int\limits_{t_1^{min}}^{t_1^{max}}dt_1
\int\limits_{z^{min}}^{z^{max}}dz
\frac{d^3\sigma^{r}(v_{cut},v_{min})}{dvdt_1dz}
\label{rad}  
\ea
is the radiative part of the cross section.
The limits of the integration over $z$ are defined from the
equation $\Delta (k_1,k_2,p_1,k)=0$, while
$t_1^{min}$ and $t_1^{max}$ can be found from $z^{min}=z^{max}$
(see \cite{zyk} for details).

Since the observable cross section (\ref{st}) is the sum of 
the radiative and non-radiative parts
\ba
\sigma^{obs}(v_{cut})
=
\sigma^{r}(v_{cut},v_{min})
+
\sigma^{nr}(v_{cut},v_{min}),
\label{st2}
\ea
we immediately find that 
\ba
\sigma^{nr}(v_{cut},v_{min})
&=&\sigma^0+\sigma^{RV}(v_{cut})+
\frac{\alpha^3}{4s}\int\limits_{0}^{v_{min}}
dv\int d\Gamma _k (|M_R|^2-4F^{IR}|M_0|^2) -
\nonumber \\&&
-
\frac{\alpha^3}{s}\int\limits_{v_{min}}^{v_{cut}}
dv\int d\Gamma _k F^{IR}|M_0|^2.
\label{norad}  
\ea

Let us notice that the explicit formulae both
 for $d^3\sigma^{r}(v_{cut},v_{min})/dvdt_1dz$ 
 in the equation (\ref{rad}) and  $\sigma^{nr}(v_{cut},v_{min})$
allow us to start the generation of the radiative events. 
However to speed up the process of generation it is 
useful to perform the integration over $z$ and $t_1$
analytically. So, the following 
analytical expressions
\ba
\frac{d\sigma^{r}(v_{cut},v_{min})}{dv}
&=&\int\limits_{t_1^{min}}^{t_1^{max}}dt_1
\int\limits_{z^{min}}^{z^{max}}dz
\frac{d^3\sigma^{r}(v_{cut},v_{min})}{dvdt_1dz},
\nonumber \\
\frac{d^2\sigma^{r}(v_{cut},v_{min})}{dvdt_1}
&=&\int\limits_{z^{min}}^{z^{max}}dz
\frac{d^3\sigma^{r}(v_{cut},v_{min})}{dvdt_1dz}
\label{rad2}  
\ea 
are incorporated in our Monte Carlo program.
The analytical integration can be performed in a standard way 
(see, for example, \cite{suarez,zyk} and references therein).

At the end of this section it should be noted that 
according to the equation (\ref{st2}) the non-radiative and radiative
contributions to the cross section depend on 
$v_{cut}$ and $v_{min}$, while the observable cross section
depends on $v_{cut}$ only.

\section{The structure of the program and 
radiative event simulation}
\label{sstruct}
\begin{figure}[t]
\unitlength 1mm
\begin{picture}(80,80)
\put(-5,-40){
\epsfxsize=15cm
\epsfysize=19cm
\epsfbox{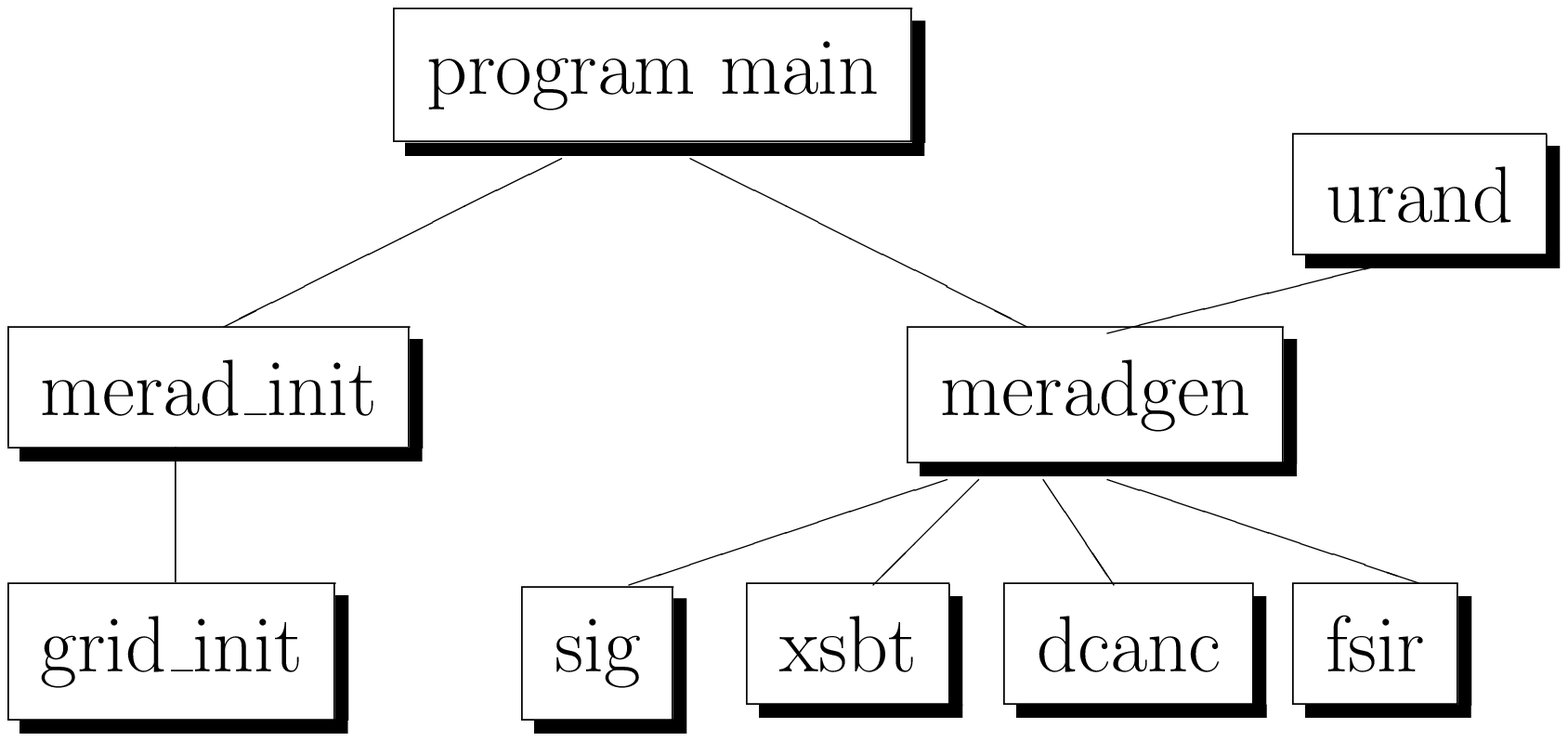}
}
\end{picture}
\vspace*{-20mm}
\caption{\label{str}
The structure of the program MERADGEN 1.0
}
\end{figure}

The structure of the Monte Carlo generator MERADGEN 1.0
 is presented in fig.~\ref{str}. 
The main blocks mean: 
\begin{itemize}
\item
{ \bf merad$\_$init } --- here we define all constants
which are necessary for generation;
\item
{ \bf grid$\_$init } --- here we prepare the grids for generation of
kinematic variables, really we approximate the theoretical
curve by some sets of segments; 
\item
{ \bf urand } --- random number generator (flat);
\item
{ \bf sig } --- the Born cross section and part of the virtual contribution
(vertices and self energies); 
\item
{ \bf xsbt } --- box contribution;
\item
{ \bf dcanc } --- the contribution with cancellation of the infrared divergency;
\item
{ \bf fsir } --- analytical cross sections
 $d\sigma^r (v_{cut},v_{min})/dv$, $d^2\sigma^r (v_{cut},v_{min})/dvdt_1$ and 
$d^3\sigma^r (v_{cut},v_{min})/dvdt_1dz$.
\end{itemize}

For the convenience sake we split our programm into 4 FORTRAN-files:

\begin{itemize}
\item
{\bf run.f} --- main program for the event generation;
\item
{\bf test.f} --- main program for the tests;
\item
{\bf meradgen.f} --- the collection of main functions and subroutines of MERADGEN 1.0; 
\item
{\bf fsir.f} --- function $f\!sir$ for calculation of the analytical cross sections that are
presented above. 
\end{itemize}
For the events (tests) generation we need to run ''make'' (''make test'') command.

Now let us consider the input-output data in more details.

\section{Input-output data} 
As an input data MERADGEN 1.0 uses 4-momentum of the virtual 
photon $vpgen:=k_1-k_2$ that generated in CM system externally,
energy of electrons and degree of electrons polarization.
There is only one variable $itest$ in MERADGEN 1.0 that responsible for output.
If $itest:=0$ the output data are gathered in two common blocks 
(see file {\bf output.inc}):

     common/variables/vgen,t1gen,zgen,weight,ich

and

      common/vectors/vprad,phirad.

Here $vgen$, $t1gen$ and $zgen$ are generated photonic 
variables $v$, $t_1$ and $z$ respectively,
(they are necessary first of all for the test), 
$weight$ is a ratio of the observable cross section to 
the Born one, variable $ich$ shows the radiative ($ich:=1$) 
or non-radiative ($ich:=0$) scattering channel, 
the 4-momentum $vprad:=p_2-p_1$ and photonic 4-momentum $phirad:=k$
also defined in CM system. 

Here we have to do some remarks:
1) for non-radiative events $vgen:=0$, $t1gen:=t$,
$z:=0$, $vprad:=vpgen$ and $phirad:=0$, and
2) as it was mentioned above, the variable $v$ can be reconstructed 
experimentally
and the events with hard photons $v_{cut}< v \leq v_{max}$
usually remove from the data. In order to speed up the
process generation, we generated variable $v$ from $v_{min}$
up to $v_{cut}=(s+t)/2 \sim v_{max}/2$.

\begin{figure}[t!]
\unitlength 1mm
\vspace*{-15mm}
\begin{tabular}{cc}
\begin{picture}(80,80)
\put(-5,0){
\epsfxsize=7cm
\epsfysize=7cm
\epsfbox{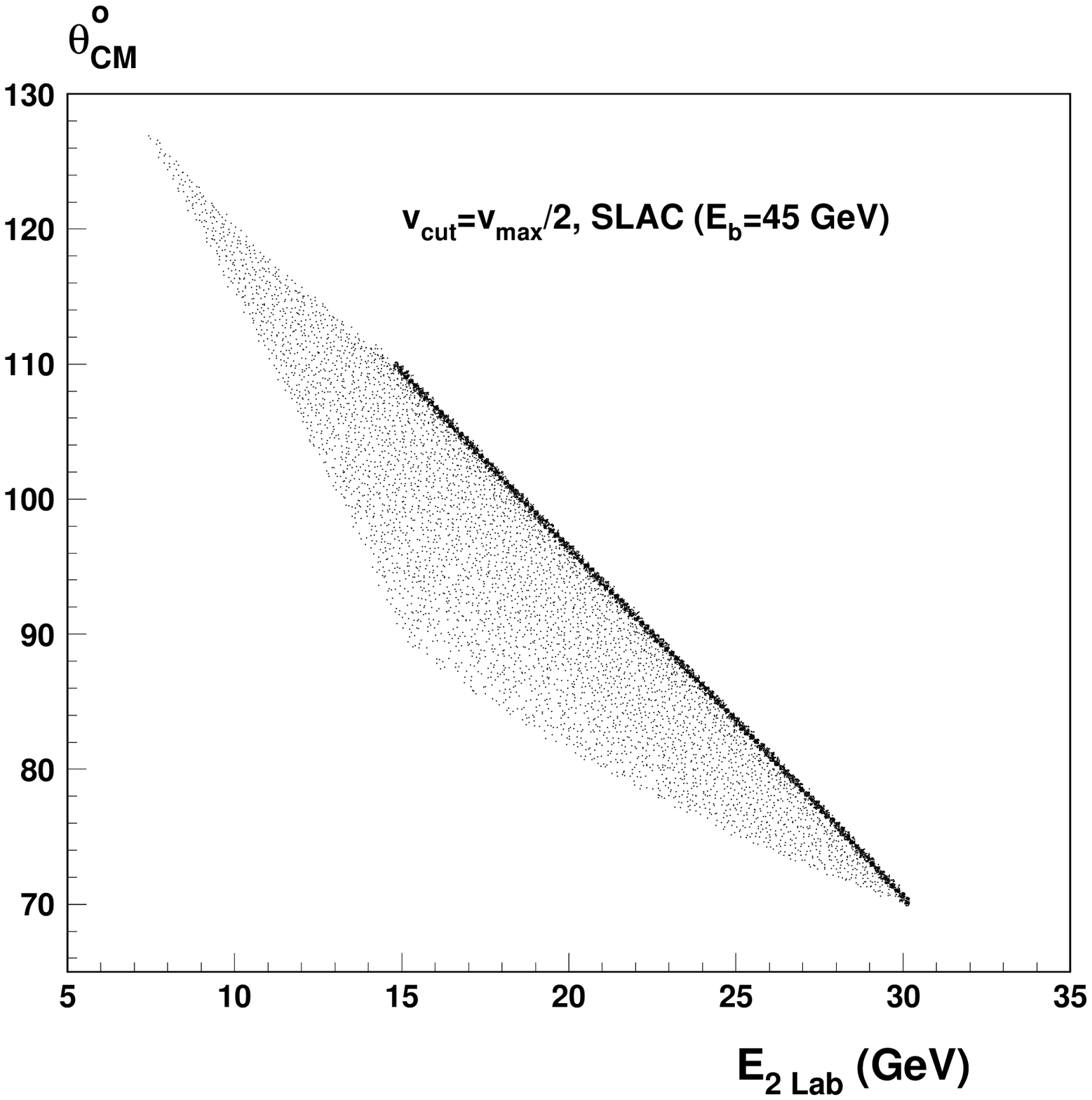}
}
\end{picture}
&
\begin{picture}(80,80)
\put(-18,0){
\epsfxsize=7cm
\epsfysize=7cm
\epsfbox{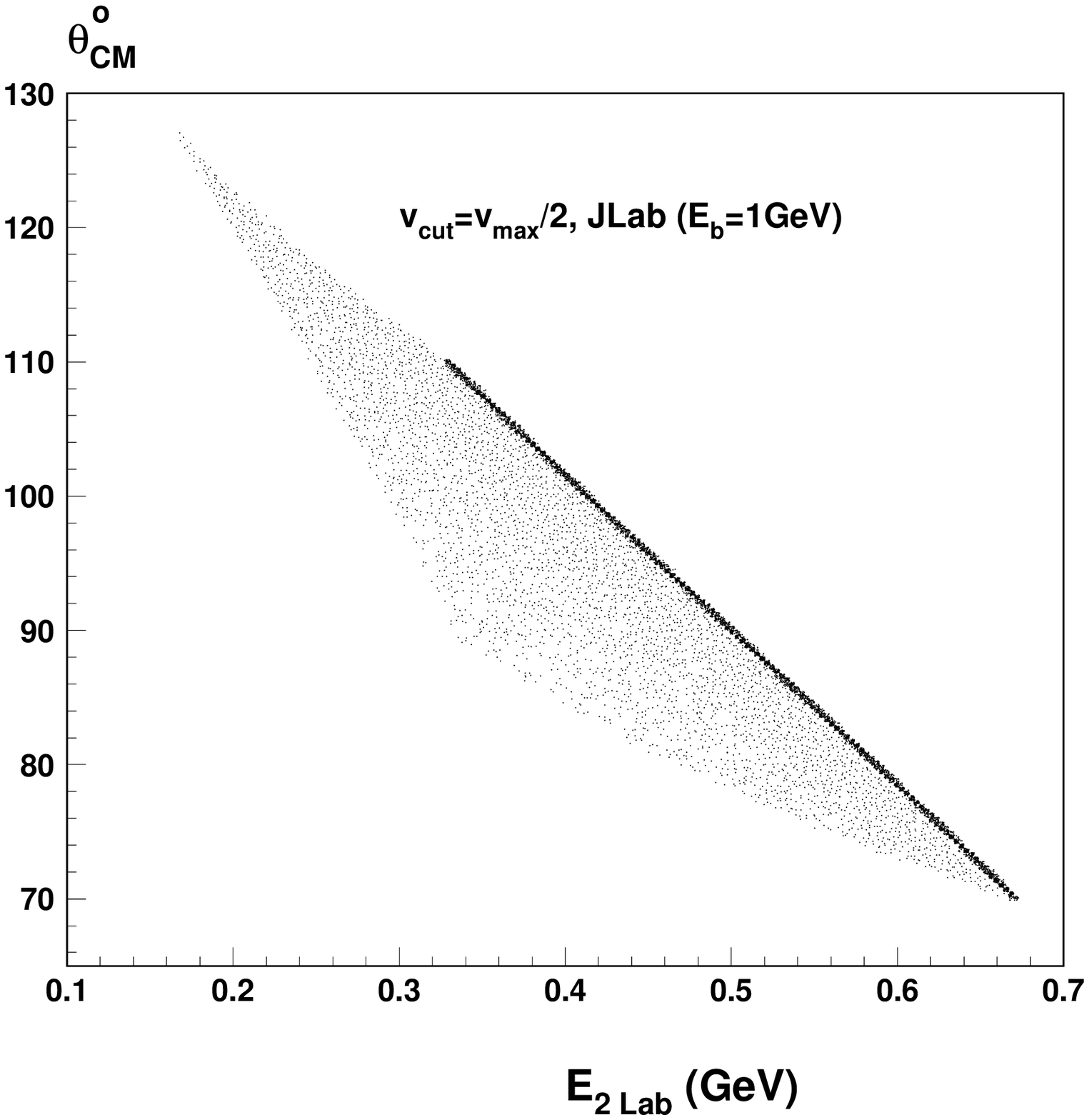}
}
\end{picture}
\end{tabular}
\caption{\label{ect}
The 2-dimensional distributions of the scattering electron energy in 
lab. system and scattering angle $\theta $ in CM system}
\end{figure}

Now let us consider one sample of generation. It is well known that
for the elastic process (\ref{el})
the energy of the detected electron $k_{20}$ can be directly defined
via the scattering angle $\theta _{\rm CM}$ and the initial electron beam 
energy in the following way:
\ba
E_2^{\rm Lab} \approx \cos ^2 \frac{\theta _{\rm CM}}{2} E_b^{\rm Lab},
\label{el2}
\ea
but when we deal with the real photon emission  (\ref{in}) 
the process becomes inelastic and, as a result, the "elastic" equation
(\ref{el2}) is broken and we have the 2-dimensional distribution
over  $E_2^{\rm Lab}$ and $\cos \theta _{\rm CM}$.
From fig.~\ref{ect} one can see that this distribution has a sharp
peak near an "elastic" line described by the equation (\ref{el2}).

\section{Test runs}  
\label{ttest}
For our test runs we use the fact that
if the events are simulated correctly their distributions over 
variables $v$, $t_1$ and $z$ must obey to the corresponding
probability distributions: 
\ba
\rho(v)&=&\frac 1{N_v}\frac {d\sigma^{r}(v_{cut},v_{min})}{dv},
\qquad
\qquad \ \ \ \ \!
N_v=\sigma^{r}(v_{cut},v_{min}),
\qquad
\nonumber\\
\rho(t_1)&=&\frac 1{N_{t_1}}\frac {d^2\sigma^{r}(v_{cut},v_{min})}{dvdt_1},
\qquad
\qquad
\;N_{t_1}=\frac {d\sigma^{r}(v_{cut},v_{min})}{dv},
\;\;\;
\nonumber\\
\rho(z)&=&\frac 1{N_z}\frac {d^3\sigma^{r}(v_{cut},v_{min})}{dvdt_1dz},
\qquad
\qquad
\;\; \ \!
N_z=\frac {d^2\sigma^{r}(v_{cut},v_{min})}{dvdt_1}.
\;\;
\label{roo}
\ea
Then for generation of $\rho(v)$, $\rho(t_1)$ or $\rho(z)$ distributions 
one has to put in the file {\bf test.f} the value of variable $itest$ such as:
$itest:=1$, $itest:=2$ or $itest:=3$, respectively, 
next to type "make test" and, at last, "./test.exe".
The value $rgen/rcalc$, i.e. ratio of generated $\rho$-distribution to 
corresponding calculated cross section should be near unit.
In \ref{tout} the test outputs for $v, t_1, z$ generation with
$P=1$ (see formula (B.2)), $E_b^{\rm Lab}$=45 GeV, $\theta_{CM}=90^0$, 
20 bins for the histogramming and $10^8$ radiative events
are presented.

\begin{figure}[ht!]
\unitlength 1mm
\begin{tabular}{cc}
\begin{picture}(80,80)
\put(-5,0){
\epsfxsize=7cm
\epsfysize=7cm
\epsfbox{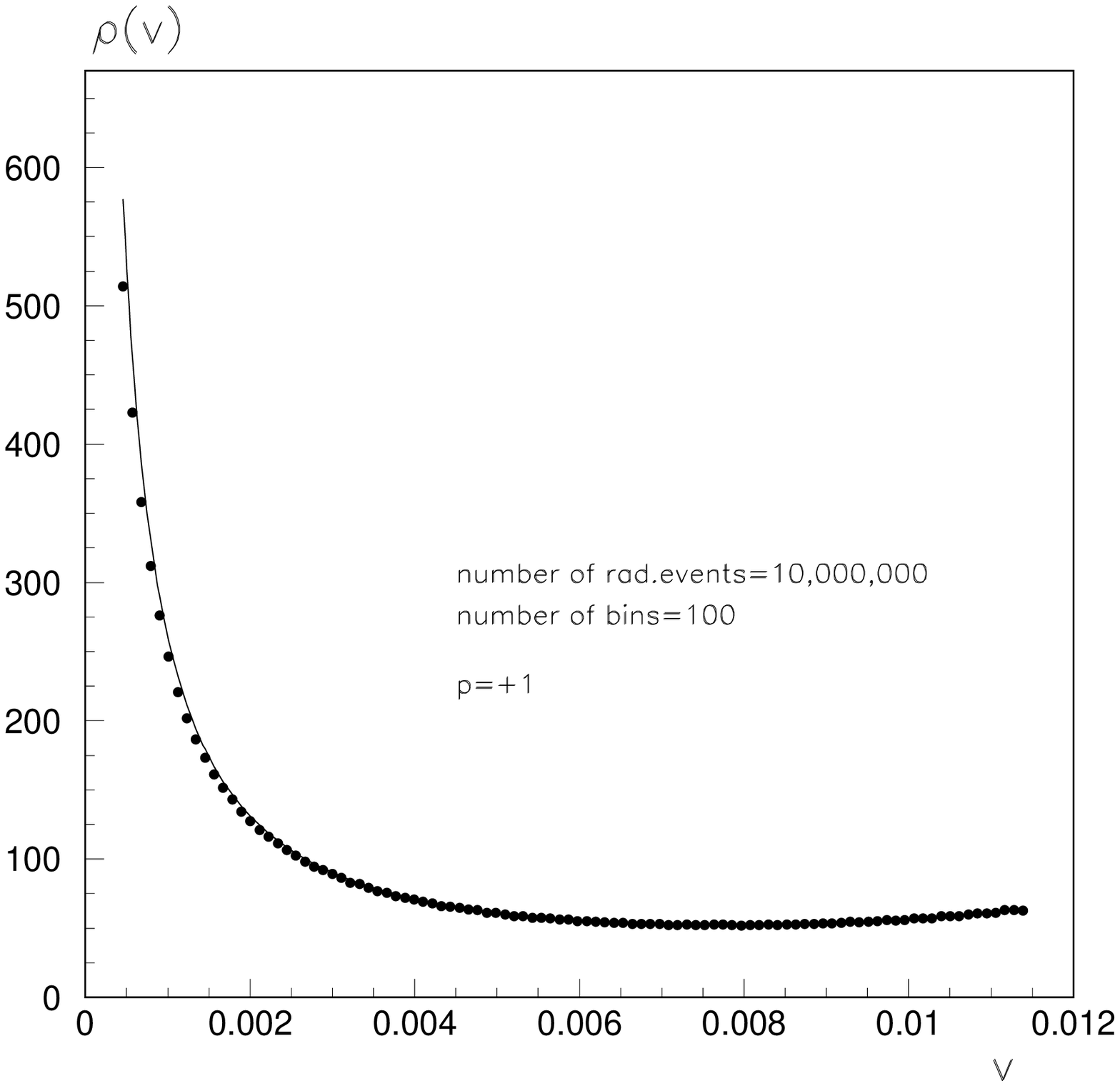}
}
\end{picture}
&
\begin{picture}(80,80)
\put(-10,0){
\epsfxsize=7cm
\epsfysize=7cm
\epsfbox{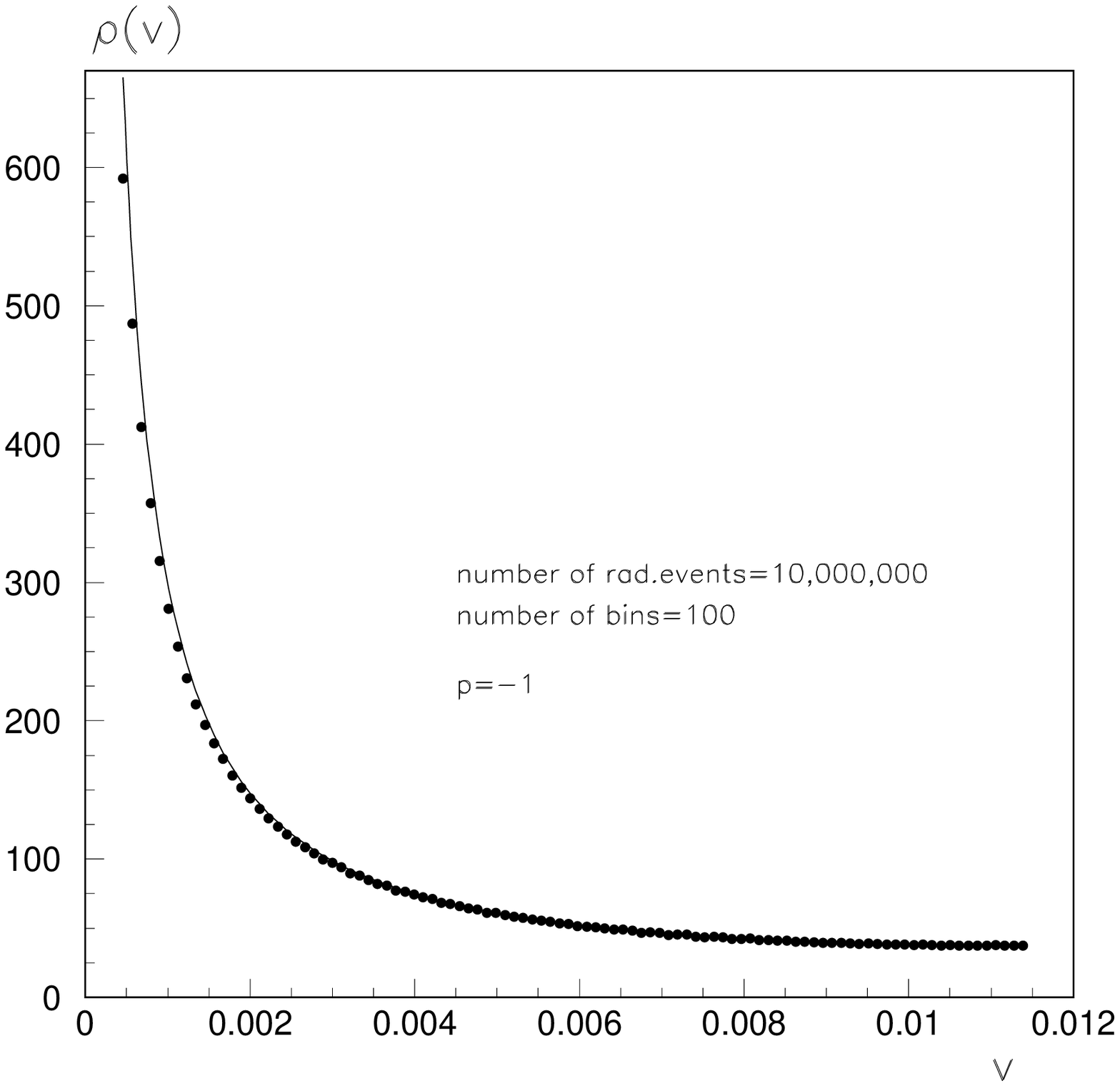}
}
\end{picture}
\\[-20mm]
\begin{picture}(80,80)
\put(-5,0){
\epsfxsize=7cm
\epsfysize=7cm
\epsfbox{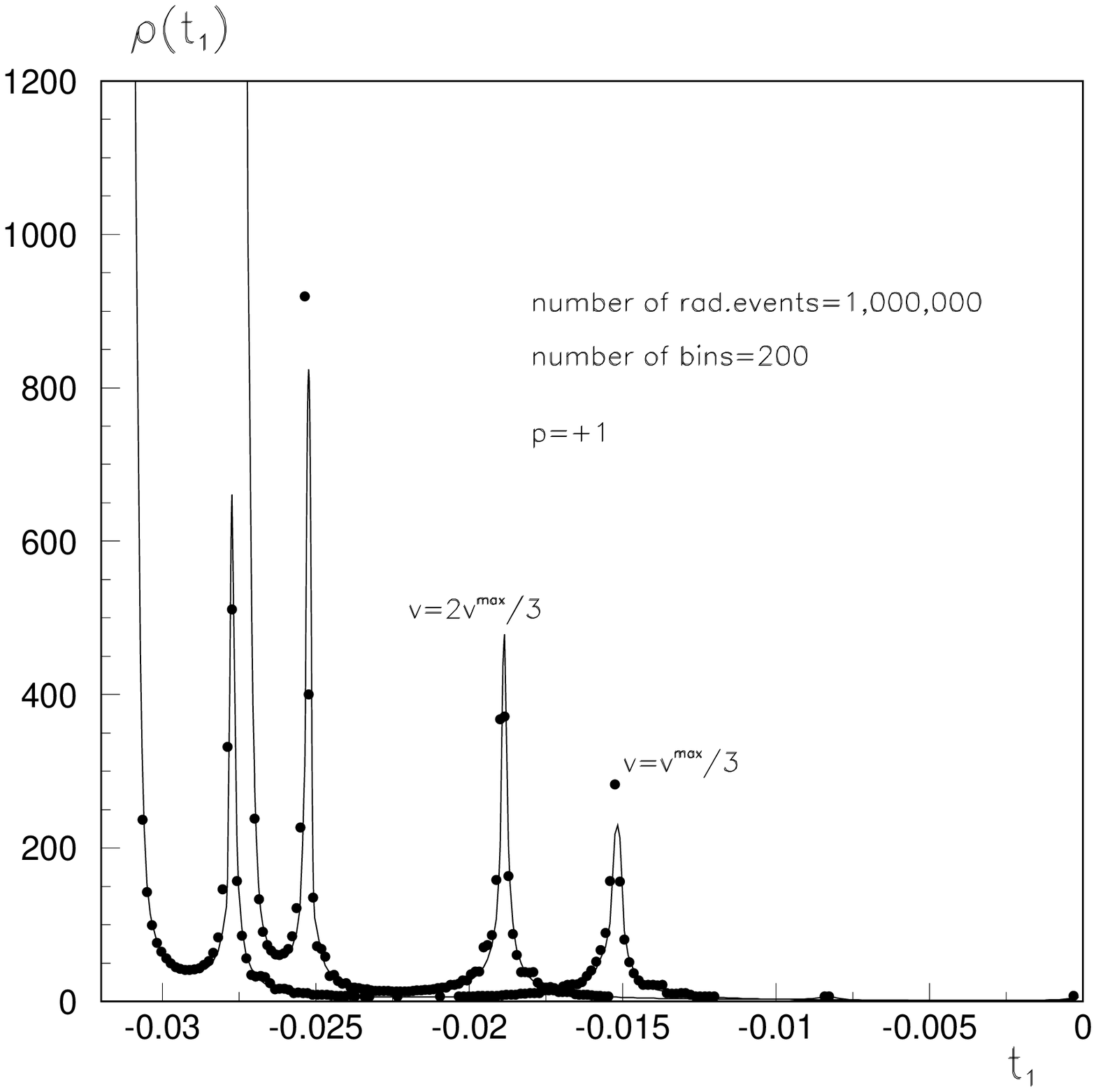}
}
\end{picture}
&
\begin{picture}(80,80)
\put(-10,0){
\epsfxsize=7cm
\epsfysize=7cm
\epsfbox{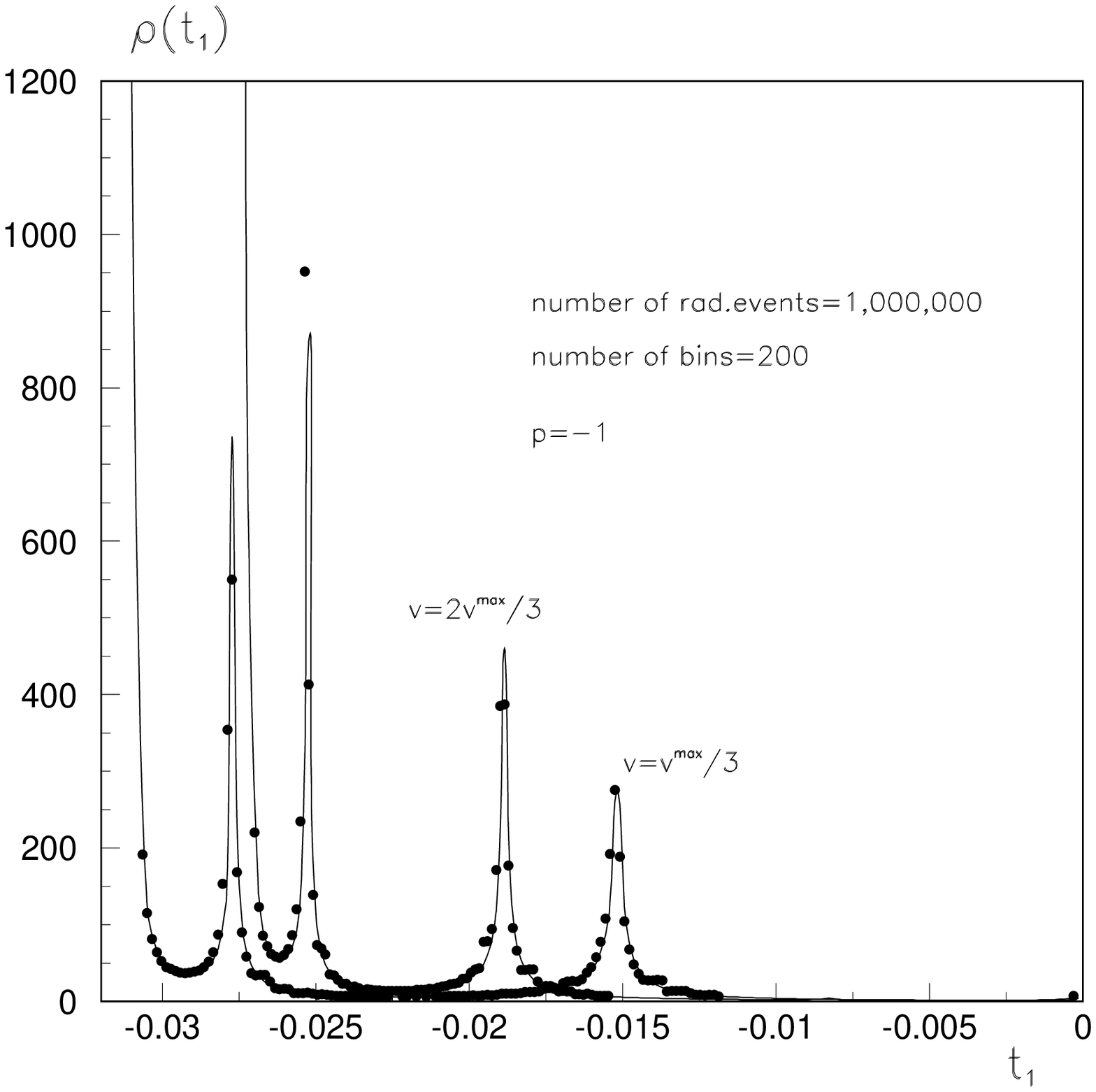}
}
\end{picture}
\\[-20mm]
\begin{picture}(80,80)
\put(-5,0){
\epsfxsize=7cm
\epsfysize=7cm
\epsfbox{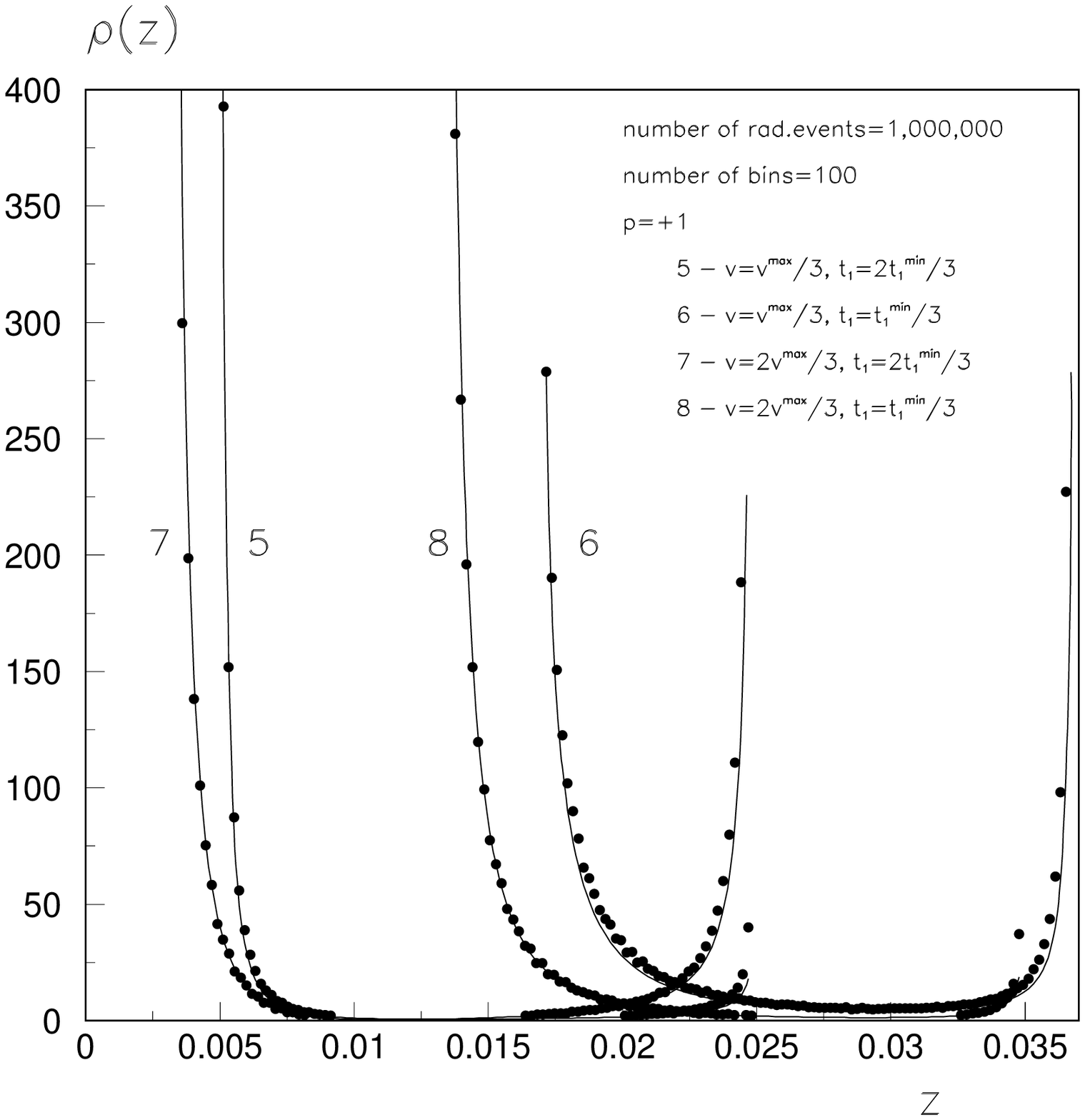}
}
\end{picture}
&
\begin{picture}(80,80)
\put(-10,0){
\epsfxsize=7cm
\epsfysize=7cm
\epsfbox{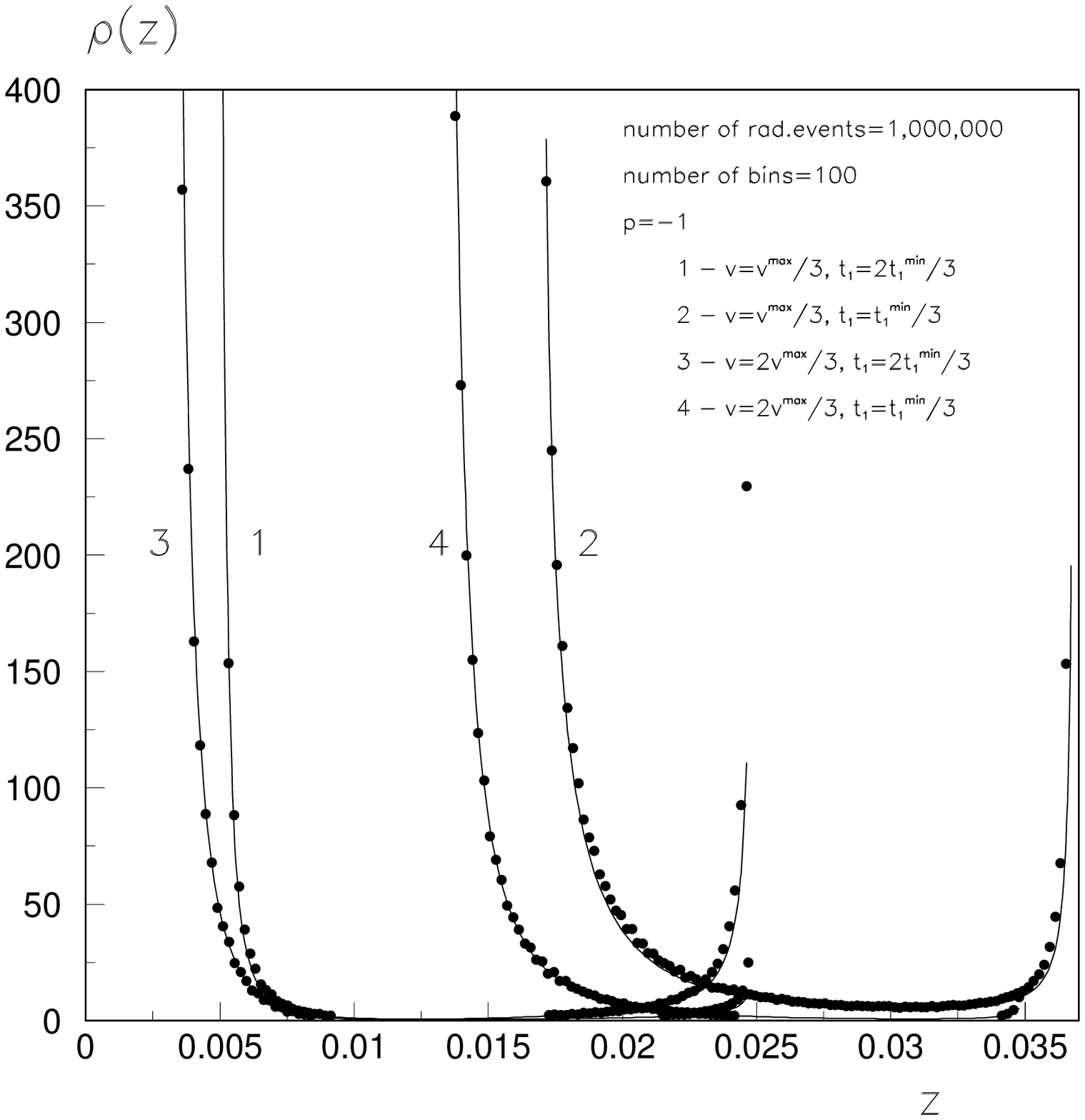}
}
\end{picture}
\end{tabular}
\vspace*{-10mm}
\caption{\label{fig3}
$v$, $t_1$, $z$-histograms (points)
and corresponding probability densities (curves)
for E158 (SLAC) kinematic conditions ($E_{b}^{\rm Lab}=45$ GeV) 
\vspace*{9mm}
}
\end{figure}

The simulated distributions of the photonic variables
for the SLAC E158 experiment kinematic conditions and 
for the different degrees of polarization are presented in fig.~\ref{fig3}
(all of parameters are noted there). 
We suppose $v_{min}$ rather small: $v_{min}=2 \times 10^{-2} E_b^{\rm Lab} m$.
In the fig.~\ref{fig3} we can see clearly the divergent
behavior of the distributions at 
$v \rightarrow 0,\ z \rightarrow z^{min, max}, t_1 \rightarrow t_1^{min}$
corresponding to the infrared singularity at the $v_{min} \rightarrow 0$.
Also it can be seen the physical, 
\newpage
so-called $z$- and $z_1$-peaks 
of distribution $\rho(t_1)$.

The other test with $itest:=4$ consists in the cross-check 
of the accuracy of vector reconstruction. So, using all 4-momenta 
we reproduce the value of generated invariant and compare 
them with the generated value.
In this test we also calculate the photonic mass square as 
$m2gamma:=phirad(4)^2-phirad(1)^2-phirad(2)^2-phirad(3)^2$,
which should be near zero.  
In last part of \ref{tout} we can see good coincidence
reconstructed and generated invariants in different (random)
kinematical points and very small values of photonic mass square.

\section{Conclusion}  
\label{cconc}
\vspace*{-5mm}
In this paper the Monte Carlo generator MERADGEN 1.0 
serving for the simulation of radiative events in 
parity conserving longitudinally doubly-polarized M{\o}ller scattering
is presented.
Following for the absolute necessity of both accuracy and quickness
for our program we have developed the fast and high precise code
using analytical integration wherever it was possible.
MERADGEN 1.0 can be employed for the radiative corrections procedure
in experiments with the complex detector geometry, such as
SLAC E158 experiment and experiments of modern polarimetry (JLAB, SLAC).

\section*{Acknowledgments}  
The authors would like to thank  Igor Akushevich,
Yury Kolomensky, Nikolai Shumeiko and Juan Suarez  
for stimulating discussions. 
V.Z. (A.I.) would like to thank SLAC (JLab) staff for 
their generous hospitality during their visits.

\begin{figure}[t!]
\unitlength 1mm
\vspace*{5mm}
\begin{picture}(80,80)
\put(25,0){
\epsfxsize=8cm
\epsfysize=8cm
\epsfbox{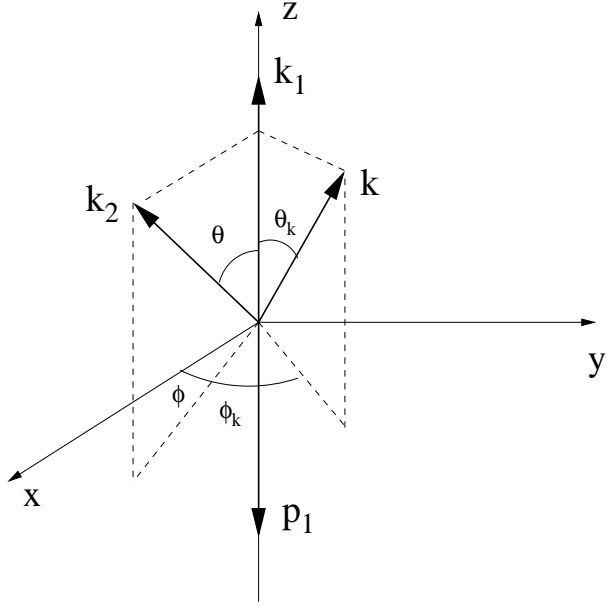}
}
\end{picture}
\vspace*{3mm}
\caption{\label{vect}
Definition of momenta and angles in center-of-mass frame}
\vspace*{3mm}
\end{figure}
\appendix
\renewcommand{\thesection}{Appendix \Alph{section}}
\renewcommand{\thesubsection}{\Alph{section}.\arabic{subsection}}
\renewcommand{\theequation}{\Alph{section}.\arabic{equation}}
\setcounter{equation}{0}

\section{4-momenta reconstruction}
\label{av}
The 4-momentum definition in the center of mass system of the initial electrons 
for M\o ller process with real photon emission is shown in fig.~\ref{vect} 
and can be presented in the form: 
\ba
&&k_1=(k_{10},0,0,|\vec{k}_{1}|),\;
p_1=(p_{10},0,0,-|\vec{p}_{1}|),
\nonumber \\
&&k_2=(k_{20},k_{21},k_{22},k_{23}),\;
p_2=(p_{20},p_{21},p_{22},p_{23}),
\nonumber \\
&&k=(k_0,k_1,k_2,k_3),
\ea
\noindent
while their components can  be expressed via the invariants and azimuthal angle
$\phi$ (that is usually generated uniformly) 
in 
the following way:
\ba
&&k_{10}=p_{10}=\frac 12 \sqrt{s},
\;|\vec{k}_{1}|=
|\vec{p}_{1}|=\frac{\sqrt{\lambda _s}}{2\sqrt{s}},
\nonumber \\
&&k_{20}=\frac{s-v}{2\sqrt{s}},\;
k_{21}=\sqrt{\frac{\lambda _3}{\lambda _s}}\cos \phi,\;
k_{22}=\sqrt{\frac{\lambda _3}{\lambda _s}}\sin \phi,\;
k_{23}=\frac{\sqrt{s}\lambda _2}{2\sqrt{\lambda _s}},
\nonumber \\
&&p_{20}=\frac{s-z}{2\sqrt{s}},
\nonumber \\
&&p_{21}=-\frac{\sqrt{\lambda_s\lambda_1\lambda_8 }\sin \phi
+(4 \lambda_3\lambda_4+s\lambda_2\lambda_7)\cos \phi
}{4\lambda_1\sqrt{\lambda_s\lambda_3}},
\nonumber \\
&&p_{22}=\frac{\sqrt{\lambda_s\lambda_1\lambda_8 }\cos \phi
-(4 \lambda_3\lambda_4+s\lambda_2\lambda_7)\sin \phi
}{4\lambda_1\sqrt{\lambda_s\lambda_3}},
\nonumber \\
&&p_{23}=\frac{\sqrt{s}(\lambda_7-\lambda_2\lambda_4)
}{2\lambda_1\sqrt{\lambda_s}},
\nonumber \\
&&k_0=\frac{v+z}{2\sqrt{s}},
\nonumber \\
&&k_1=\frac{\sqrt{\lambda _s\lambda _1\lambda _8}\sin \phi 
+(4\lambda _3\lambda _6-s\lambda _2\lambda _7)\cos \phi }
{4\lambda _1\sqrt{\lambda _s\lambda _3}},
\nonumber \\
&&k_2=\frac{-\sqrt{\lambda _s\lambda _1\lambda _8}\cos \phi 
+(4\lambda _3\lambda _6-s\lambda _2\lambda _7)\sin \phi }
{4\lambda _1\sqrt{\lambda _s\lambda _3}},
\nonumber \\
&&k_3=\frac{\sqrt{s}(\lambda_7+ \lambda_2 \lambda_6)}{2\lambda_1 \sqrt{\lambda_s} }. 
\ea
Here 
\ba
&&\lambda _1=(s-v)^2-4sm^2,
\qquad
\qquad
\qquad
\lambda _2=2t+s-v-4m^2,
\nonumber \\&&
\lambda _3=-st(s+t-v-4m^2)-m^2v^2,\;
\lambda _4=s(s-v-4m^2)-(s+v)z,\;
\nonumber \\&&
\lambda _5=vz(s-v-z)-m^2(v+z)^2,
\;\;\;\;
\lambda _6=s(v-z)-v(v+z),\;
\nonumber \\&&
 \lambda _7=(s+2t_1-z-4m^2)\lambda_1-\lambda_2\lambda_4,\;
\lambda_8=16\lambda_3\lambda_5-\lambda_7^2.
\ea

As a result, the angles in fig.~\ref{vect}
can be expressed via the invariants in the following way:
\ba
\cos \theta=\frac{s\lambda _2}{\sqrt{\lambda _s\lambda _1}},
\;
\cos \theta_k=\frac{s(\lambda_7+ \lambda_2 \lambda_6)}
{(v+z)\lambda_1 \sqrt{\lambda_s} },
\;
\tan \phi_k=\frac{\sqrt{\lambda _s\lambda _1\lambda _8}}
{s\lambda _2\lambda _7-4\lambda _3\lambda _6}.
\ea

\section{Additional virtual particle, $\delta_1^H$ and $\delta_1^S$ contributions}
\label{avc}
The virtual contributions to M{\o}ller scattering
can be separated into
three parts:
\begin{equation}
\sigma^{\rm V}=
\sigma^{\rm S}+\sigma^{\rm Ver}+\sigma^{\rm Box},
\label{G}
\end{equation}
where 
1) $\sigma^{\rm S}$ is a virtual photon self-energy contribution,
2) $\sigma^{\rm Ver}$ is a vertex function  contribution,
3) $\sigma^{\rm Box}$ is a box  contribution. Now we consider each of them.
\begin{enumerate}
\item
The contribution of the virtual photon self energies 
(including the photon vacuum polarization by hadrons)
to the cross section looks like
\begin{eqnarray}
\sigma^{\rm S}&=& \frac{4\pi \alpha^2}{t^2}
\mbox{Re} \Bigl( -\frac{1}{t} \hat{\Sigma}^{\gamma}_T(t)
 + \Pi_h(-t) \Bigr )
\Bigl[ 
 (1+P)\frac{u^2}{s}
   -(1-P)\frac{s^2}{u}  \Bigr ]
\nonumber \\[0.3cm] &&
 + (t \leftrightarrow u).
\label{self-en}
\end{eqnarray}
Here  $P=P_BP_T$, where $P_B$ ($P_T$) is the beam (target) polarization,    
$\hat{\Sigma}^{\gamma}_T(-t)$ is the renormalized transverse
part of the $\gamma$--self-energy \cite{BSH86}
(this part includes
vacuum polarization by  $e$, $\mu$ and $\tau$ charged leptons:
in corresponding formula of \cite{BSH86} we should take a
summing index $f=e,\mu,\tau$).
The hadronic part of the photonic vacuum polarization associated with
light quarks can be directly obtained from the data on
process $e^+e^- \rightarrow \mbox{hadrons}$ via dispersion relations.
Here we use parameterization of \cite{Bur-Piet}
\begin{equation}
\mbox{Re} \Pi_h(-t) \cong A+B \ln (1+C|t|),
\end{equation}
with updated parameters A,B,C in different energy regions.

\item
For the contribution of the electron vertices we used the
results of the paper \cite{BSH86} (see also references therein).
We can obtain the vertex part as
\begin{eqnarray}
\sigma^{\rm Ver}&=&\frac{2 \alpha^3}{t^2}
\Bigl[  (1+P)\frac{u^2}{s}
   -(1-P)\frac{s^2}{u}  \Bigr] \Lambda_1(t,m^2)
 + (t \leftrightarrow u),
\label{csv}
\end{eqnarray}
where
\begin{eqnarray}
\Lambda_1(t,m^2) = -2\ln\frac{|t|}{\lambda^2} 
\Bigl(\ln\frac{|t|}{m^2}-1\Bigr)
+ \ln\frac{|t|}{m^2} + \ln^2\frac{|t|}{m^2} + \frac{\pi^2}{3}-4.
\end{eqnarray}
\item
Recalculated here expressions for the box cross section 
are slightly different from presented in \cite{mera}
(we correct the misprints in the expression (16) of
\cite{mera}) 
\begin{eqnarray}
\sigma^{\rm Box}&=&
\frac{2 \alpha^3}{t}
\Bigl[   \frac{1+P}{s}
\Bigl( \frac{2u^2}{t}\ln\frac{s}{|u|}\ln\frac{
\sqrt{s|u|}}{\lambda^2}
-\delta^1_{(\gamma\gamma)} \Bigr)
\nonumber \\[0.3cm] \displaystyle&& 
-  \frac{1-P}{u}
       \Bigl( \frac{2s^2}{t}\ln\frac{s}{|u|}\ln\frac{
\sqrt{s|u|}}{\lambda^2}
         -\delta^2_{(\gamma\gamma)} \Bigr ) \Bigr ]
 + (t \leftrightarrow u),
\end{eqnarray}
The expressions $\delta_{(\gamma\gamma)}^{1,2}$ have the form:
\begin{eqnarray}
\delta^1_{(\gamma \gamma)} & = &
l_s^2\frac{  s^2+u^2}{2t} - l_su -(l_x^2+\pi^2)\frac{u^2}{t},
\nonumber \\[0.3cm] \displaystyle
\delta^2_{(\gamma \gamma)} &=&
  l_s^2\frac{s^2}{t} + l_x s - (l_x^2 + \pi^2 )\frac{s^2+u^2}{2t},
\end{eqnarray}
and logarithms look like
\begin{equation}
l_s=\ln\frac{s}{|t|},\ l_x=\ln\frac{u}{t}.
\end{equation}
\end{enumerate}
It should be noted that vertex and box parts contain the infrared divergence
through the appearance of the fictitious photon mass $\lambda$.
The infrared part from virtual cross section
can be extracted in a simple way:
\begin{eqnarray}
\sigma^{V}_{IR} &&=
\sigma^{V} - \sigma^{V}(\lambda^2 \rightarrow s)
= -\frac{2\alpha}{\pi} \ln\frac{s}{\lambda^2}
  \Bigl( \ln\frac{tu}{m^2s}-1 \Bigr) \sigma^{0}.
\end{eqnarray}
The correct expressions for $\delta_1^H$ and  $\delta_1^S$
read
\ba
\delta_1^H  & = &-\frac 1 2l_m^2+
\Bigl (\ln \frac{t^2(s+t)^2(s-v_{max})}{s(s+t-v_{max})^2v_{max}(v_{max}-t)}
+1\Bigr )l_m
-\frac 1 2 \ln ^2\frac{v_{max}}{|t|} -
\nonumber \\[0.3cm] && 
-\ln ^2\bigl(1-\frac{v_{max}}{t}\bigr)
+\ln \frac {s+t}{s+t-v_{max}}
\ln \frac {(s+t)(s+t-v_{max})}{t^2} +
\nonumber \\[0.3cm]&&
+\ln \frac {s-v_{max}}{|t|}
\ln \frac {s-v_{max}}{s}
+\ln \frac {v_{max}}{|t|}
+2\Bigl[
{\rm Li}_2\left(\frac{v_{max}}s\right)
-{\rm Li}_2\left(\frac{v_{max}}t\right) -
\nonumber \\[0.2cm]&&
-{\rm Li}_2\left(\frac{v_{max}}{s+t}\right)\Bigl]
+{\rm Li}_2\left(\frac{s-v_{max}}s\right)
-{\rm Li}_2\left(\frac{t-v_{max}}t\right)
-\frac{\pi^2}6,
\nonumber \\[0.3cm]
\delta_1^S &=& -\frac 5 2l_m^2+(3-2 l_r)l_m
-(l_m-1)\ln \frac{s(s+t)}{t^2}-\frac 12l_r^2
-\frac{\pi ^2}{3}+1.
\ea

\section{Matrix element of the real photon emission}
\label{arc}
The sum of the matrix elements contributed to the real 
photon emission in M{\o}ller process reads:
\begin{eqnarray}
M_R^{\alpha}&=&
\frac 1{t_1}
\bar{u}(k_2)\Gamma _{\mu \alpha }(k_2,k_1)u(k_1)
\bar{u}(p_2)\gamma _{\mu}u(p_1) +
\nonumber\\
&+& \frac 1{t}
\bar{u}(k_2)\gamma _{\mu}u(k_1)
\bar{u}(p_2)\Gamma _{\mu \alpha }(p_2,p_1)u(p_1) -
\nonumber\\
&-&
\frac 1{u}
\bar{u}(p_2)\Gamma _{\mu \alpha }(p_2,k_1)u(k_1)
\bar{u}(k_2)\gamma _{\mu}u(p_1) -
\nonumber\\
&-&
\frac 1{z_2}
\bar{u}(p_2)\gamma _{\mu}u(k_1)
\bar{u}(k_2)\Gamma _{\mu \alpha }(k_2,p_1)u(p_1).
\end{eqnarray}
The conjugate matrix element can be found in a simple way:
\begin{eqnarray}
{\bar M}_R^{\alpha}&=&
\frac 1{t_1}
\bar{u}(k_1)\bar{\Gamma} _{\nu \alpha }(k_1,k_2)u(k_2)
\bar{u}(p_1)\gamma _{\nu}u(p_2) +
\nonumber\\&&
+
\frac 1{t}
\bar{u}(k_1)\gamma _{\nu}u(k_2)
\bar{u}(p_1)\bar{\Gamma }_{\nu \alpha }(p_1,p_2)u(p_2) -
\nonumber\\&&
-
\frac 1{u}
\bar{u}(k_1)\bar{\Gamma }_{\nu \alpha }(k_1,p_2)u(p_2)
\bar{u}(p_1)\gamma _{\nu}u(k_2) -
\nonumber\\&&
-
\frac 1{z_2}
\bar{u}(k_1)\gamma _{\nu}u(p_2)
\bar{u}(p_1)\bar{\Gamma }_{\nu \alpha }(p_1,k_2)u(k_2).
\end{eqnarray}
Here $z_2=z-s-t_1+4m^2$ and
\begin{eqnarray}
\Gamma _{\mu \alpha}(a,b)=
\Bigl(\frac {b_{\alpha }}{kb}-\frac {a_{\alpha }}{ka}
\Bigr)\gamma _{\mu}
-\frac{\gamma _{\mu}\hat{k}\gamma _{\alpha}}{2 bk}
-\frac{\gamma _{\alpha }\hat{k}\gamma _{\mu}}{2 ak},\;
\nonumber\\
\bar{\Gamma }_{\nu \alpha}(a,b)=
-\Bigl(\frac {b_{\alpha }}{kb}-\frac {a_{\alpha }}{ka}
\Bigr)\gamma _{\nu}
-\frac{\gamma _{\nu}\hat{k}\gamma _{\alpha}}{2 bk}
-\frac{\gamma _{\alpha }\hat{k}\gamma _{\nu}}{2 ak}.
\end{eqnarray}
By introducing
\begin{eqnarray}
&&S(a_1,a_2,a_3,a_4)={\rm Tr}[a_1\rho (a_2)a_3\Lambda (a_4)]
\nonumber\\
&&S(a_1,a_2,a_3,a_4,a_5,a_6,a_7,a_8)=
{\rm Tr}[a_1\rho (a_2)a_3\Lambda (a_4)a_5\rho (a_6)a_7\Lambda (a_8)],
\end{eqnarray}
where
\begin{eqnarray}
&&u(k_1)\bar{u}(k_1)=\rho(k_1)=
\frac 12(\hat{k_1}+m)(1-P_B\gamma _5\hat{\xi}_L),\;
\nonumber \\
&&u(p_1)\bar{u}(p_1)=\rho(p_1)=
\frac 12(\hat{p_1}+m)(1-P_T\gamma _5\hat{\eta}_L),
\nonumber \\
&&u(k_2)\bar{u}(k_2)=\Lambda(k_2)=
\hat{k}_2+m,\ \
u(p_2)\bar{u}(p_2)=\Lambda(p_2),
\end{eqnarray}
the square of matrix elements reads:
\begin{eqnarray}
|M_R|^2 &=& -M_R^{\alpha } {\bar M}_R^{\alpha } =
\nonumber\\
&=&
-\frac 1{t_1^2}
S(\Gamma _{\mu \alpha} (k_2, k_1),k_1,\bar {\Gamma} _{\nu \alpha
}(k_1,k_2),k_2)
S(\gamma _{\mu},p_1,\gamma _{\nu },p_2) -
\nonumber\\&&
-\frac 1{t_1t}
S(
\Gamma _{\mu \alpha} (k_2, k_1),k_1,
\gamma _{\nu }
,k_2)
S(\gamma _{\mu},p_1,
\bar {\Gamma} _{\nu \alpha }(p_1,p_2)
,p_2) -
\nonumber\\&&
-\frac 1{t_1t}
S(
\gamma _{\mu }
,k_1,
\bar{\Gamma }_{\nu \alpha} (k_1, k_2)
,k_2)
S(
\Gamma _{\mu \alpha }(p_2,p_1)
,p_1,
\gamma _{\nu}
,p_2) -
\nonumber\\&&
-\frac 1{t^2}
S(
\gamma _{\mu }
,k_1,
\gamma _{\nu}
,k_2)
S(
\Gamma _{\mu \alpha }(p_2,p_1)
,p_1,
\bar{\Gamma }_{\nu \alpha} (p_1, p_2)
,p_2) +
\nonumber\\&&
+\frac 1{t_1u}
S(
\Gamma _{\mu \alpha }(k_2,k_1)
,k_1,
\bar{\Gamma }_{\nu \alpha} (k_1, p_2)
,p_2,
\gamma _{\mu}
,p_1,
\gamma _{\nu}
,k_2
) +
\nonumber\\&&
+\frac 1{t_1u}
S(
\Gamma _{\mu \alpha }(p_2,k_1)
,k_1,
\bar{\Gamma }_{\nu \alpha} (k_1, k_2)
,k_2,
\gamma _{\mu}
,p_1,
\gamma _{\nu}
,p_2
) +
\nonumber\\&&
+\frac 1{t_1z_2}
S(
\Gamma _{\mu \alpha }(k_2,k_1)
,k_1,
\gamma _{\nu}
,p_2,
\gamma _{\mu}
,p_1,
\bar{\Gamma }_{\nu \alpha} (p_1, k_2)
,k_2
) +
\nonumber\\&&
+\frac 1{t_1z_2}
S(
\gamma _{\mu}
,k_1,
\bar{\Gamma }_{\nu \alpha} (k_1, k_2)
,k_2,
\Gamma _{\mu \alpha }(k_2,p_1)
,p_1,
\gamma _{\nu}
,p_2
) +
\nonumber\\&&
+\frac 1{tu}
S(
\gamma _{\mu}
,k_1,
\bar{\Gamma }_{\nu \alpha} (k_1, p_2)
,p_2,
\Gamma _{\mu \alpha }(p_2,p_1)
,p_1,
\gamma _{\nu}
,k_2
) +
\nonumber\\&&
+\frac 1{tu}
S(
\Gamma _{\mu \alpha }(p_2,k_1)
,k_1,
\gamma _{\nu}
,k_2,
\gamma _{\mu}
,p_1,
\bar{\Gamma }_{\nu \alpha} (p_1, p_2)
,p_2
) +
\nonumber\\&&
+\frac 1{tz_2}
S(
\gamma _{\mu}
,k_1,
\gamma _{\nu}
,p_2,
\Gamma _{\mu \alpha }(p_2,p_1)
,p_1,
\bar{\Gamma }_{\nu \alpha} (p_1, k_2)
,k_2
) +
\nonumber\\&&
+\frac 1{t_1z_2}
S(
\gamma _{\mu}
,k_1,
\gamma _{\nu}
,k_2,
\Gamma _{\mu \alpha }(k_2,p_1)
,p_1,
\bar{\Gamma }_{\nu \alpha} (p_1, p_2)
,p_2
)  -
\nonumber\\&&
-\frac 1{u^2}
S(\Gamma _{\mu \alpha} (p_2, k_1),k_1,\bar {\Gamma} _{\nu \alpha
}(k_1,p_2),p_2)
S(\gamma _{\mu},p_1,\gamma _{\nu },k_2) -
\nonumber\\&&
-\frac 1{uz_2}
S(
\Gamma _{\mu \alpha} (p_2, k_1),k_1,
\gamma _{\nu }
,p_2)
S(\gamma _{\mu},p_1,
\bar {\Gamma} _{\nu \alpha }(p_1,k_2)
,k_2) -
\nonumber\\&&
-\frac 1{uz_2}
S(
\gamma _{\mu }
,k_1,
\bar{\Gamma }_{\nu \alpha} (k_1, p_2)
,p_2)
S(
\Gamma _{\mu \alpha }(k_2,p_1)
,p_1,
\gamma _{\nu}
,k_2) -
\nonumber\\&&
-\frac 1{z_2^2}
S(
\gamma _{\mu }
,k_1,
\gamma _{\nu}
,p_2)
S(
\Gamma _{\mu \alpha }(k_2,p_1)
,p_1,
\bar{\Gamma }_{\nu \alpha} (p_1, k_2)
,k_2).
\end{eqnarray}
\section{Test output}
\label{tout}
Here we present the results of the test as {\bf test.dat} output file
corresponding to:\\
1) $itest:=1$ -- the generation of $\rho(v)$ distribution 
and comparison it with the analytical cross section
corresponding to the first formula in (\ref{roo})
(here and below all of invariants $v, t_1, z$ are in $\mbox{GeV}^2$)
\input{testv.dat}
2) $itest:=2$ -- the generation of $\rho(t_1)$ distribution 
and comparison it with the analytical cross section
corresponding to the second formula in (\ref{roo})
\input{testt1.dat}
3) $itest:=3$ -- the generation of $\rho(z)$ distribution 
and comparison it with the analytical cross section
corresponding to the third formula in (\ref{roo})
\input{testz.dat}
4) $itest:=4$ -- the cross-check 
of the accuracy of the vector reconstruction for 5 random radiative events 
\input{testvec.dat}

\end{document}